%% file: main.tex
\documentclass[10pt,twocolumn,letterpaper]{article}

\usepackage[pagenumbers]{cvpr}

\input{preamble}
\usepackage{array}
\usepackage{dblfloatfix}

\definecolor{cvprblue}{rgb}{0.21,0.49,0.74}
\usepackage[breaklinks,colorlinks,allcolors=cvprblue]{hyperref}

\hypersetup{
  pdftitle={TetherMark: Instance-Level Watermark Authentication Against Cross-Instance Transplantation},
  pdfauthor={Jin Wei, Rundong Li, Li Ma, Yikai Wang, Ruihao Yang, Jianxiong Wu, Xiaoyuan Duan, Yanbo Wang, Runyu Mao, Chen Fang, Chang Xu, Ping Chen, Jun Dai, Xiaoyan Sun}
}

\title{\method: Instance-Level Watermark Authentication Against Cross-Instance Transplantation}

\author{
{\normalsize Jin Wei$^{1}$, Rundong Li$^{2}$, Li Ma$^{1}$, Yikai Wang$^{1}$, Ruihao Yang$^{1}$}\\
{\normalsize Jianxiong Wu$^{3}$, Xiaoyuan Duan$^{2}$, Yanbo Wang$^{2}$, Runyu Mao$^{1}$, Chen Fang$^{1}$}\\
{\normalsize Chang Xu$^{1}$, Ping Chen$^{1,\dagger}$, Jun Dai$^{4,\dagger}$, Xiaoyan Sun$^{4}$}\\[0.45em]
{\small $^{1}$Institute of Big Data, Fudan University, Shanghai, China}\\
{\small $^{2}$School of Computing, Xi'an Jiaotong-Liverpool University}\\
{\small $^{3}$School of Computer Science, Fudan University, Shanghai, China}\\
{\small $^{4}$Department of Computer Science, Worcester Polytechnic Institute, MA, USA}\\[-0.1em]
{\small $^{\dagger}$Corresponding authors.}
}

\begin{document}
\maketitle
\begin{abstract}
Robust watermarking is commonly verified by asking whether a hidden payload remains decodable after compression, geometric distortion, or generative editing. Localized and semantic extensions may additionally check whether an expected region or category is still present. These tests establish evidence validity or content consistency, but not whether the evidence still belongs to the particular visual instance registered by its owner. We expose \emph{cross-instance transplantation}: an attacker extracts or estimates valid watermark evidence from one registered image and transfers it to a different image of the same category. The message can still decode, the expected content can remain present, and the result can look natural, even though the evidence has been reassigned to the wrong instance. To enable \emph{instance-level watermark authentication}, we introduce \method, a post-hoc watermark that combines two complementary checks. The message check verifies an edit-robust payload, while the relation check uses secret patch-pair selection, a registered response, natural correspondence transport, and enrollment normalization to test consistency with the registered instance. The verifier accepts only when both checks pass. Across 240 test subjects, \method{} attains 92.08\% legitimate true acceptance with 9.17\% and 16.25\% attack success under WmForger and WMCopier, respectively, yielding the highest instance-binding balanced accuracy under both attacks.
\end{abstract}

\section{Introduction}
\label{sec:intro}

Most watermark-based provenance systems frame verification as a robustness test: can the registered payload still be recovered after the image has been transformed? Traditional detectors therefore make a message-level decision. If the registered bits survive compression, geometric distortion, or generative editing, the query is accepted as watermarked~\cite{bui2025trustmark,hu2024robustwide,lu2025vine,soucek2025pixelseal}. This confirms that watermark evidence is recoverable, but it does not compare the query with the visual instance for which that evidence was issued.

Content-aware watermarking brings verification closer to the protected content, but usually at a coarser level. Localized schemes check whether a designated region remains, while semantic approaches check whether the query still contains the expected category, caption, or concept~\cite{sander2025wam,liu2025wofa,arabi2025seal,zhang2026sembind}. A query can therefore pass once it still contains, for example, a watch in the protected region; the check need not establish that it is the \emph{same watch}. Payload validity and category/region consistency are thus distinct from the finer-grained question of whether valid evidence still belongs to the enrolled instance.

Consider a marketplace that registers a photograph of one specific wristwatch. An attacker estimates the watermark residual from that photograph and places it on a different wristwatch of the same product category. The traditional detector recovers the exact registered message and accepts. A semantic or localized auxiliary verifier also sees a watch in the expected subject region and accepts. The forged image can remain visually natural, yet the registered evidence now accompanies the wrong physical item. We call this failure \emph{cross-instance transplantation} and study its hardest same-category case. Our objective is \emph{instance-level watermark authentication}: accept legitimate edits of the registered watch while rejecting a different watch with copied evidence.

To this end, we introduce \method, an instance-level authentication framework that explicitly separates evidence validity from registered-instance consistency. The message check verifies that at least one robust payload branch carries the registered message. The relation check verifies a private enrollment relation: the key selects patch pairs inside the subject, enrollment records their response, natural correspondences transport the same pairs through an edit, and enrollment normalization tests whether the aligned response remains consistent with the registered instance. The final decision is an AND rule. Thus a copied message may pass the first check, while the enrollment-relation check rejects its reassignment to a different instance.

Unlike Rel-Zero~\cite{chen2026relzero}, which uses intrinsic relations as a
stand-alone zero-watermark, \method{} makes a keyed, enrollment-relative
relation the registered-instance check of a post-hoc watermark authenticator.
Across 240 test subjects and 2{,}400 legitimate edits, \method{} attains a
92.08\% true acceptance rate (TAR) and limits same-category attack success rate
(ASR) to 9.17\%/16.25\% under WmForger/WMCopier, yielding the highest
instance-binding balanced accuracies (91.46\%/87.92\%). Under 1{,}000-step
key-unknown transfer, ASR remains 8.33\%; on unseen categories, TAR is 84--92\% with 4\%
copy ASR. Together, these results make message validity and relation consistency
necessary for instance-level acceptance.

Our contributions are threefold:
\begin{itemize}
    \item We formulate cross-instance transplantation as a distinct threat and evaluation target: valid payload and category evidence must not authenticate the wrong instance.
    \item We propose an instance-level watermark authenticator that accepts only when a robust message and a keyed, enrollment-relative subject relation are both valid; natural correspondence preserves it across legitimate edits.
    \item We define a calibrated verification protocol and evaluate the mechanism
    with public-partition, no-shaping, enrollment-normalization, key-unknown
    transfer, unseen-category, and cost controls.
\end{itemize}

\section{Related Work}
\label{sec:related_work}

\paragraph{Robust payload watermarking.}
Classical methods embed spread-spectrum or perceptually masked transform-domain
signals~\cite{cox1997spread,barni2001wavelet}, while learned encoder--decoders
train payload recovery under distortions, physical recapture, unknown
corruptions, adversarial training, and JPEG compression
~\cite{zhu2018hidden,tancik2020stegastamp,luo2020distortion,wen2019romark,
jia2021mbrs}. Diffusion-era schemes root evidence in model parameters or
generative latents~\cite{fernandez2023stable,wen2023treering,
yang2024gaussian,ci2024ringid,zhang2024zodiac,xian2024raw}. Post-hoc systems
such as TrustMark, Robust-Wide, VINE, and Pixel Seal further improve recovery
under arbitrary-resolution, instruction-driven, diffusion-based, or
high-resolution edits~\cite{bui2025trustmark,hu2024robustwide,lu2025vine,
soucek2025pixelseal}. Their verifier remains message-centric: a decoded payload
establishes evidence presence, while its assignment to the registered image
instance remains unchecked.

\paragraph{Localized and semantic binding.}
Content-aware methods more closely associate evidence with what remains in the
query. WAM localizes evidence and WOFA distributes it against partial theft
~\cite{sander2025wam,liu2025wofa}; SWIFT, SEAL, and SemBind bind evidence to
captions or semantics~\cite{evennou2024swift,arabi2025seal,zhang2026sembind}.
MetaSeal uses content-dependent cryptographic watermarks
~\cite{zhou2025metaseal}, while Zhu et al. target content watermarking against
removal and forgery~\cite{zhu2026ists}. These designs verify stronger content consistency than a
global payload alone. A different instance can still share the same category,
protected region, caption, or prompt. Instance-level authentication therefore
requires a registered-instance consistency test.

\paragraph{Watermark forgery and transplantation.}
Recent attacks separate detector acceptance from legitimate provenance.
Black-box attacks transfer semantic evidence across diffusion
generators~\cite{muller2025blackbox}; WmForger optimizes from one marked
example~\cite{soucek2025wmforger}; WMCopier injects a learned marked-image
signal through shallow inversion~\cite{dong2025wmcopier}; and WForge recovers
transferable watermark-like residuals~\cite{wang2026wforge}. WAVES standardizes
distortion, regeneration, and adversarial tests~\cite{an2024waves}, while
regeneration and averaging expose removal and reusable-pattern failure modes
~\cite{zhao2024removable,yang2024averaging}. These studies establish that
watermark evidence can be forged, transferred, or removed. Their success
criteria measure evidence manipulation broadly; our evaluation isolates the
same-category reassignment of a valid payload while expected content remains
present.

\paragraph{Relational authentication.}
Rel-Zero~\cite{chen2026relzero} registers naturally stable patch-pair relations
as a zero-watermark for generative editing. \method{} extends relational
evidence with a conventional robust payload and a separately evaluated
registered-instance check. Secret pair selection and enrollment normalization
form the relation check, correspondence transports it to the query, and the AND
rule combines it with message validity. Active shaping further separates the
registered response.

\begin{figure*}[t]
    \centering
    \includegraphics[width=\textwidth]{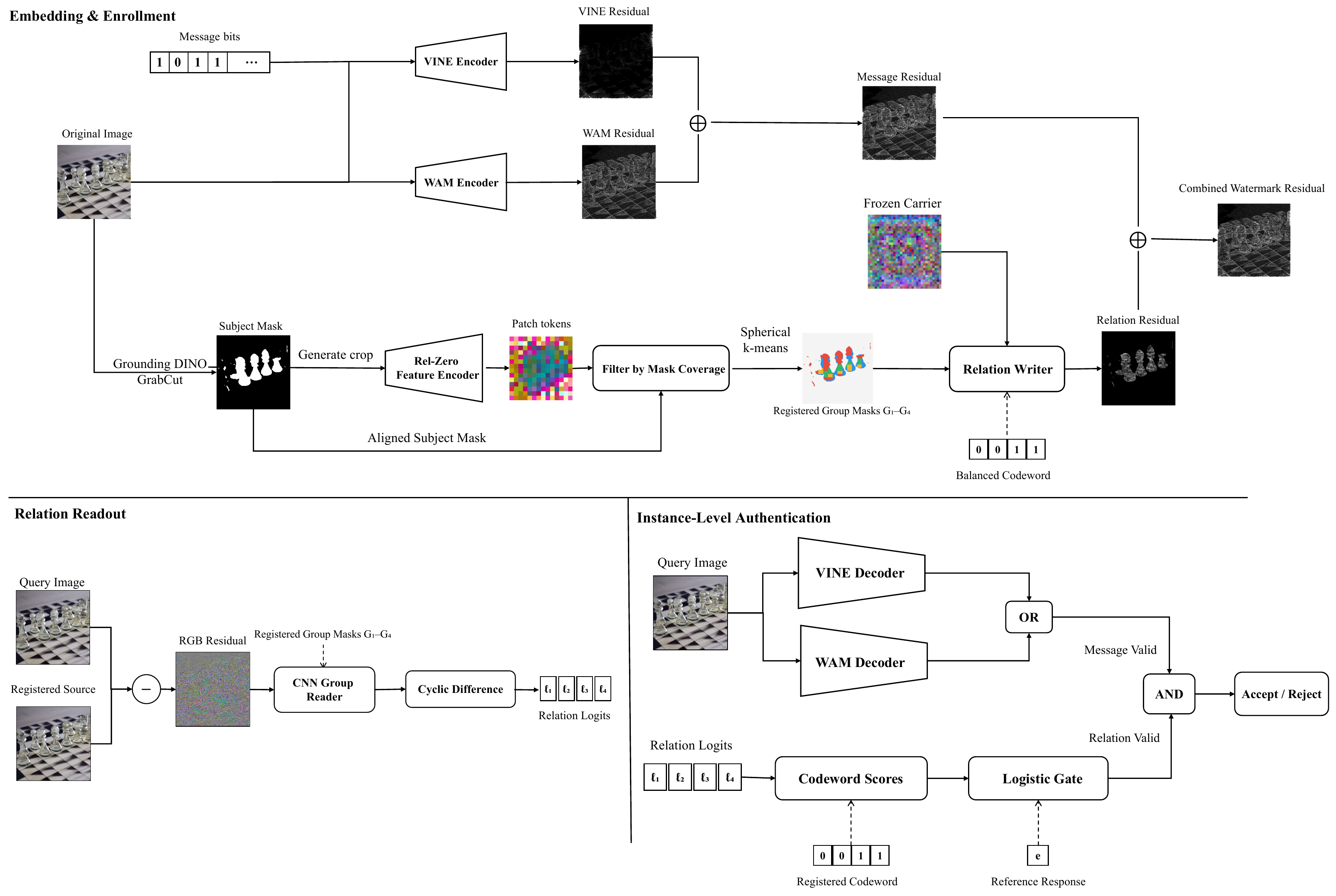}
    \caption{Overview of the \method{} workflow. \textbf{Embedding and
    enrollment.} VINE and WAM residuals form the message carrier. Mask-filtered
    Rel-Zero tokens define four registered groups; a fixed carrier, the keyed
    balanced codeword, and the relation writer produce the relation residual.
    \textbf{Relation readout.} After source--query alignment, the RGB residual
    and registered group masks pass through the CNN group reader and cyclic
    difference to obtain relation logits. \textbf{Instance-level
    authentication.} An OR rule combines the two payload decoders, while
    codeword scores and the enrollment reference feed the relation gate. An
    AND rule accepts only when both paths are valid.}
    \label{fig:workflow}
\end{figure*}

\section{Problem Formulation and Threat Model}
\label{sec:problem}

\paragraph{Enrollment and legitimate transformations.}
The owner holds an image $x\in[0,1]^{H\times W\times 3}$, a subject mask $M$, a payload $m$, and a secret key $k$. Enrollment produces a marked image $x_w$ and a private record $r$ describing the protected instance. Given a query image $y$, the verifier receives $(y,m,k,r)$ and must determine whether $y$ is a legitimate transformation of the enrolled instance. We allow changes to the background, illumination, material, style, viewpoint, and moderate generative appearance, provided that provenance continuity of the enrolled subject is preserved. Thus, a stylized rendering of an enrolled watch may be accepted, whereas replacing it with an unrelated watch should not be, even if both images remain semantically consistent. The private record contains the registered source-subject cache and all information needed for verification; the edit prompt and diffusion seed remain outside the interface.

\paragraph{Cross-instance transplantation.}
In contrast to a legitimate transformation, a transplantation attack extracts or estimates valid watermark evidence from registered instance $A$ and places it on a different instance $B$. The adversary knows the algorithms and public feature extractors, obtains at least one marked image, and may copy or estimate its watermark residual. The adversary can apply generative editing and optimize a bounded perturbation against a public identity matcher. Our primary case chooses $B$ from the same semantic category as $A$: message validity, category consistency, and visual naturalness can all be preserved while the evidence--instance assignment is wrong. The key $k$ and enrollment record $r$ remain secret. We additionally test key-unknown transfer from white-box optimization against a public subject verifier.

\paragraph{Instance-level authentication objective.}
The message check asks whether registered watermark evidence is present; the
relation check asks whether that evidence remains consistent with the enrolled
instance. The final decision is
\begin{equation}
\label{eq:accept}
\operatorname{Accept}(y)
=
\operatorname{MessageValid}(y,m,k)
\wedge
\operatorname{RelationValid}(y,r,k).
\end{equation}
Our primary evaluation targets same-category cross-instance transplantation,
where payload and semantic checks remain plausible. Unmarked images, wrong
keys, and subject replacement form auxiliary rejection cases around this
instance-assignment test.

\section{Method}
\label{sec:method}

\method{} implements Eq.~\eqref{eq:accept} through the three stages in
Fig.~\ref{fig:workflow}. Embedding and enrollment fuse the VINE/WAM message
carrier with a keyed group-relation carrier and save the private record.
Relation readout aligns the registered source and group masks with a query,
then converts their RGB residual into relation logits. Instance-level
authentication evaluates the decoder OR path and relation-gate path in
parallel, followed by the final AND decision.

\subsection{Embedding and Enrollment}
\label{sec:embedding_enrollment}

The Embedding and Enrollment stage in Fig.~\ref{fig:workflow} takes image $x$,
subject mask $M$, payload $m$, and secret key $k$. VINE and WAM encode their
registered branch payloads and their residuals are summed into the message carrier.
Inside $M$, frozen Rel-Zero tokens are filtered by mask coverage and clustered
into four registered group masks. The key selects ordered group/patch pairs and
a balanced codeword; starting from a fixed carrier, the relation writer shapes
the corresponding low-amplitude state. Enrollment then records the response
observed in the released image, the group geometry, and correspondence data.
Thus this stage releases $x_w$ and registers $r$.

Let $\delta_{\mathrm{V}}$ and $\delta_{\mathrm{W}}$ be the VINE and WAM
residuals for $(x,m,k)$, respectively. For a candidate relation perturbation
$\delta$, define
\begin{equation}
\label{eq:hybrid_carrier}
\begin{aligned}
\delta_{\mathrm{msg}}&=\delta_{\mathrm{V}}+\delta_{\mathrm{W}},\\
z(\delta)&=\operatorname{clip}(x+\delta_{\mathrm{msg}}+M\odot\delta,0,1),\\
x_w&=z(\delta^*),\qquad \delta_{\mathrm{rel}}=M\odot\delta^*.
\end{aligned}
\end{equation}
Here $\delta^*$ is obtained below. Both branches contribute to the same protected image, with the relation perturbation restricted to $M$. The message branch provides payload capacity and edit robustness. The relation branch supplies the keyed response used for instance-level consistency, while enrollment anchors that response to the registered source.

Let $\mathcal{G}_x=\{G_1,\ldots,G_4\}$ denote the registered group masks.
The secret key selects relation components and $J$ ordered group/patch pairs,
\begin{equation}
\label{eq:keyed_pairs}
\mathcal{P}_k=\{(a_j,b_j,u_j)\}_{j=1}^{J},
\end{equation}
where $a_j,b_j$ index supports in $\mathcal{G}_x$ and $u_j$ selects a component
of the shared CNN group reader. Let $d(z,x)=z-x$ denote the source-aligned RGB
residual at enrollment, and let $g(d,p)$ be the reader descriptor of residual
$d$ restricted to support $p$. The response associated with each tuple is
\begin{equation}
\label{eq:relation_response}
s_j(z,x;k)=\left\langle u_j,\,g(d(z,x),a_j)-g(d(z,x),b_j)\right\rangle.
\end{equation}
The ordered difference makes the response depend on a relation between two patches. The key also samples a balanced code $c\in\{-1,+1\}^{J}$ with $\sum_j c_j=0$. Complementary signs discourage a common-direction response shortcut. With four components, six balanced patterns are available; our evaluated one-bit instantiation uses one complementary pair and treats the other patterns as wrong-key alternatives. Pair and component choices provide further key diversity.

The fixed carrier shown in Fig.~\ref{fig:workflow} initializes $\delta$; the
encoders, Rel-Zero extractor, and relation reader remain frozen while the writer
optimizes its pixels. We write the code on the combined image $z(\delta)$:
\begin{equation}
\label{eq:relation_write}
\begin{split}
\delta^*=\argmin_{\delta}\;&
\frac{1}{J}\sum_{j=1}^{J}
\log\!\left(1+\exp[-c_j s_j(z(\delta),x;k)]\right)\\
&+\lambda_q\mathcal{L}_{\mathrm{quality}}(x,z(\delta)),\\
\text{s.t. }&\|M\odot\delta\|_{\infty}\leq\epsilon_{\mathrm{rel}}.
\end{split}
\end{equation}
The quality loss combines pixel and perceptual distortion penalties against the original image. Including $\delta_{\mathrm{msg}}$ in $z(\delta)$ makes relation shaping account for the image actually released. The budget $\epsilon_{\mathrm{rel}}$ limits the additional perturbation; payload recovery must also be checked on the resulting $x_w$ because both branches share the carrier. Section~\ref{sec:relation-ablation} separately tests whether shaping is necessary once the private pair selection, registered response, correspondence transport, and enrollment-normalized gate are retained.

After substituting $\delta^*$ into Eq.~\eqref{eq:hybrid_carrier}, we evaluate Eq.~\eqref{eq:relation_response} on $x_w$ and register
\begin{equation}
\label{eq:enrollment_record}
\begin{aligned}
e_j&=s_j(x_w,x;k),\qquad j=1,\ldots,J,\\
r&=\{\mathcal{X}_x,e,\mathcal{P}_k,\mathcal{D}_x,\mathcal{G}_x\}.
\end{aligned}
\end{equation}
Here $\mathcal{X}_x$ is the registered source-subject cache and
$e\in\mathbb{R}^{J}$ is the reference response read from the released image.
$\mathcal{D}_x$ stores frozen tokens, while $\mathcal{G}_x$ stores the four
group masks and their subject-relative geometry for alignment. The
correspondence descriptors can differ from the relation reader $g$; no paired
opposite-code counterpart is required.

\subsection{Relation Readout}
\label{sec:relation_readout}

During the Relation Readout stage in Fig.~\ref{fig:workflow}, the verifier
receives query $y$ and enrollment record $r$. Because generative editing can
move, reshape, or partially redraw a subject, we use
$(\mathcal{D}_x,\mathcal{G}_x)$ from Eq.~\eqref{eq:enrollment_record} to locate
registered patches in the query. Frozen grayscale token descriptors provide
mutual nearest-neighbor matches between the registered source and $y$. Let
$\pi_y$ map a matched source location to its query location, and let
$\widetilde{\mathcal{X}}_y$ denote the source cache transported into the query
frame. We transport the registered masks in $\mathcal{G}_x$ by the same map.
The subtraction node in Fig.~\ref{fig:workflow} forms the aligned RGB residual
$d_y=y-\widetilde{\mathcal{X}}_y$. The shared CNN group reader processes $d_y$
on each transported mask. Its cyclic-difference block implements the ordered
subtraction for every registered pair:
\begin{equation}
\label{eq:transported_response}
q_j(y;k,r)=\left\langle u_j,
g(d_y,\pi_y(a_j))-g(d_y,\pi_y(b_j))\right\rangle.
\end{equation}
Thus, the displayed relation logits $q=(q_1,\ldots,q_J)$ measure the same
ordered group relations as the enrolled response $e$. Frozen descriptor
similarity establishes correspondence; the keyed CNN readout supplies the
authentication evidence.

We denote correspondence validity by $C(y,r,k)\in\{0,1\}$. It requires reliable matches for the selected pair endpoints and sufficient subject match coverage and confidence, with thresholds fixed using only the 60 calibration subjects. If $C=0$, the query is rejected before constructing $q$ or evaluating the relation gate. This rejection rule is included explicitly in the verifier below.

\subsection{Instance-Level Authentication}
\label{sec:instance_authentication}

The Instance-Level Authentication stage in Fig.~\ref{fig:workflow} evaluates
two paths in parallel. VINE and WAM decode the query independently; the OR node
sets $\operatorname{MessageValid}=1$ when either branch passes its own frozen
threshold. We denote the aggregated result by $D_{\mathrm{msg}}$. On the
relation path, $C=0$ directly gives $\mathrm{RelationValid}=0$; for $C=1$, the
codeword-score block compares $q$ in
Eq.~\eqref{eq:transported_response} with the registered response $e$. Raw
relation magnitude varies across subjects and edit strengths, so a large
response alone does not establish agreement with enrollment. We therefore
feed both raw statistics and enrollment-relative features to a linear gate:
\begin{equation}
\label{eq:gate_features}
\begin{aligned}
\phi_{\mathrm{raw}}(y,r,k)
&=[q^{\downarrow};\mu(q);\sigma_q;\gamma_k;\gamma_{\neg k}],\\
\phi_{\mathrm{norm}}(q,e)
&=\left[\frac{q}{|e|+\varepsilon};\cos_{\varepsilon}(q,e);
\alpha_e(q);\|q-e\|_2\right],\\
\phi(y,r,k)&=[\phi_{\mathrm{raw}};\phi_{\mathrm{norm}}].
\end{aligned}
\end{equation}
Semicolons denote concatenation; $q^{\downarrow}$ is the sorted response vector, and $\mu(q),\sigma_q$ are its mean and standard deviation. The ratio is elementwise, with $\varepsilon>0$ for stability. We use $\cos_{\varepsilon}(q,e)=q^\top e/(\|q\|_2\|e\|_2+\varepsilon)$ and the projection coefficient $\alpha_e(q)=e^\top q/(\|e\|_2^2+\varepsilon)$.

The ``Codeword Scores'' block in Fig.~\ref{fig:workflow} produces
\begin{equation}
\label{eq:code_margins}
\begin{aligned}
\gamma_k&=\frac{1}{J}\sum_{j=1}^{J}c_jq_j(y;k,r),\\
\gamma_{\neg k}&=\max_{k'\in\mathcal{K}_{-}}
\frac{1}{J}\sum_{j=1}^{J}c_j^{(k')}q_j(y;k',r).
\end{aligned}
\end{equation}
Here $\mathcal{K}_{-}$ is a sampled set of wrong keys. We regenerate their codes
and pairs as in Sec.~\ref{sec:embedding_enrollment}, then recompute their
responses with Eq.~\eqref{eq:transported_response} on retained enrollment
tokens. Consequently, $\gamma_{\neg k}$ compares independently sampled
wrong-key responses. The sampling protocol is fixed before test evaluation.

The displayed ``Logistic Gate'' is a calibrated linear map from
Eq.~\eqref{eq:gate_features} to the relation score. Including correspondence
rejection gives
\begin{equation}
\label{eq:relation_probability}
B(y,r,k)=
\begin{cases}
0, & C(y,r,k)=0,\\
\operatorname{sigmoid}\!\left(\beta^\top[1;\widehat{\phi}(y,r,k)]\right),
& C(y,r,k)=1,
\end{cases}
\end{equation}
where $\widehat{\phi}$ is standardized using means and scales estimated exclusively from the 60 calibration subjects, and $\beta$ includes a bias coefficient. The gate parameters and its decision threshold are fitted only on allowed edits and cross-instance transplants from these calibration subjects. All correspondence thresholds, feature statistics, gate parameters, key-sampling protocol, and authentication thresholds are then frozen. The remaining 60 development subjects are used only to validate this frozen rule and do not contribute to feature standardization, gate fitting, or threshold selection. The final test subjects are used only for evaluation.

Finally, let $D_{\mathrm{msg}}(y,m,k)$ denote the possibly aggregated payload-branch confidence; the evaluated aggregation is specified below. For frozen thresholds $\tau_{\mathrm{msg}}$ and $0<\tau_{\mathrm{rel}}<1$, Eq.~\eqref{eq:accept} becomes
\begin{equation}
\label{eq:final_decision}
\begin{aligned}
\operatorname{Accept}(y)
={}&\mathbb{1}[D_{\mathrm{msg}}(y,m,k)\geq\tau_{\mathrm{msg}}]\\
&\cdot\mathbb{1}[B(y,r,k)\geq\tau_{\mathrm{rel}}].
\end{aligned}
\end{equation}
The first indicator is the decoder OR path and tests whether registered evidence
is present. The second is the relation-gate path, including correspondence
validity, and tests consistency with the enrolled instance. Their product is
the final AND node in Fig.~\ref{fig:workflow}. A transplanted payload can pass
the first path while failing the second. The key-unknown transfer experiment
measures this separation against optimization through a public verifier.

\subsection{Implementation Details}
\label{sec:implementation}

\paragraph{Partition and writing.}
We retain RGB within the subject mask, fill its exterior with $127/255$,
and resize the subject bounding-box crop to $224\times224$. The frozen
Rel-Zero student ViT~\cite{chen2026relzero} yields a $14\times14$ grid
of $\ell_2$-normalized tokens. We require at least eight patches with
subject occupancy of at least $0.25$. Spherical $k$-means uses farthest-point
initialization and at most 20 iterations to produce the four groups.
We cap the merged carrier at an ROI RMS of $9/255$ and an
$\ell_\infty$ norm of $32/255$, then upsample it from $256\times256$ to
$512\times512$ and apply three rounds of RMS restoration and clipping.
The fusion weight $\rho=0.25$ gives nominal relation-component bounds of
$2.25/255$ (ROI RMS) and $8/255$ ($\ell_\infty$).

\paragraph{Readout and decision.}
The shared CNN group reader applies four stride-two convolutions to each
registered-mask residual, then forms the cyclic pair differences in
Eq.~\eqref{eq:transported_response}. The convolutions use kernel sizes
$(5,3,3,3)$ and output widths $(w,2w,4w,4w)$. Each convolution is
followed by GELU. Adaptive average pooling and a linear layer
produce a scalar. We fit the relation gate with sample-weighted
logistic regression using an $\ell_2$ penalty of $1.0$ on the
coefficients and an unpenalized intercept. Newton/IRLS runs for up
to 100 iterations and stops when the largest absolute parameter
update falls below $10^{-9}$. In our redundant payload branch,
$\operatorname{MessageValid}$ holds when either at least 71 of the 100 VINE-B
bits match or all 32 WAM bits match.

% Declare the two leading result floats before the experiment text so the
% full-width robustness table can be placed within the eight-page body.
\begin{table}[t]
\centering
\footnotesize
\setlength{\tabcolsep}{2.5pt}
\caption{Same-category cross-instance transplantation. TAR pools 2{,}400 edited
queries. Each attack column reports ASR / IBA over 240 attempts. All values are
percentages.}
\label{tab:transplantation}
\begin{tabular}{@{}
>{\raggedright\arraybackslash}m{0.32\columnwidth}
>{\centering\arraybackslash}m{0.15\columnwidth}
>{\centering\arraybackslash}m{0.225\columnwidth}
>{\centering\arraybackslash}m{0.225\columnwidth}@{}}
\toprule
Method & Legit. TAR $\uparrow$
& \shortstack{WmForger\\ASR $\downarrow$ / IBA $\uparrow$}
& \shortstack{WMCopier\\ASR $\downarrow$ / IBA $\uparrow$} \\
\midrule
TrustMark-P & 89.38 & 80.00 / 54.69 & 90.83 / 49.27 \\
Pixel Seal & 89.75 & 84.58 / 52.58 & 85.42 / 52.17 \\
Robust-Wide & 80.00 & 100.00 / 40.00 & 99.17 / 40.42 \\
WAM & 77.46 & 83.33 / 47.06 & 95.83 / 40.81 \\
SWIFT & 74.38 & 77.92 / 48.23 & 89.58 / 42.40 \\
VINE & 69.96 & 84.58 / 42.69 & 97.50 / 36.23 \\
VINE + Learned Subject Check & 63.63 & 85.42 / 39.10 & 95.00 / 34.31 \\
\method{} (ours) & 92.08 & 9.17 / 91.46 & 16.25 / 87.92 \\
\bottomrule
\end{tabular}
\end{table}

\begin{table*}[t]
\centering
\small
\setlength{\tabcolsep}{3.5pt}
\caption{Robustness to legitimate edits. TAR (\%) under image processing,
compositing, and generative editing. Each condition contains 240 eligible
queries. Redrawing uses denoising strengths of 0.25 and 0.35.}
\label{tab:legitimate-edits}
\resizebox{\textwidth}{!}{%
\begin{tabular}{l*{10}{c}}
\toprule
& \multicolumn{2}{c}{Image processing} & \multicolumn{3}{c}{Compositing and background}
& \multicolumn{5}{c}{Subject appearance} \\
\cmidrule(lr){2-3}\cmidrule(lr){4-6}\cmidrule(lr){7-11}
\shortstack{Method} & \shortstack{JPEG-70} & \shortstack{$0.5\times$ scale}
       & \shortstack{Cut-and-paste} & \shortstack{Cool bg.}
       & \shortstack{Warm bg.} & \shortstack{Material}
       & \shortstack{Watercolor} & \shortstack{Technical\\style}
       & \shortstack{Redraw 0.25} & \shortstack{Redraw 0.35} \\
\midrule
TrustMark-P & 95.83 & 100.00 & 75.00 & 72.92 & 74.17 & 95.83 & 95.00 & 94.17 & 95.83 & 95.00 \\
Pixel Seal & 100.00 & 100.00 & 99.58 & 95.00 & 92.08 & 80.83 & 83.33 & 80.42 & 84.17 & 82.08 \\
Robust-Wide & 97.92 & 99.58 & 80.42 & 78.33 & 79.58 & 72.08 & 71.25 & 70.00 & 75.42 & 75.42 \\
WAM & 81.67 & 100.00 & 94.58 & 95.42 & 92.92 & 62.92 & 61.67 & 62.92 & 60.83 & 61.67 \\
SWIFT & 90.00 & 95.00 & 75.42 & 73.33 & 74.58 & 67.08 & 65.83 & 65.00 & 69.17 & 68.33 \\
VINE & 100.00 & 100.00 & 2.08 & 2.08 & 2.08 & 98.33 & 98.75 & 99.17 & 99.58 & 97.50 \\
VINE + Learned Subject Check & 99.58 & 99.58 & 2.08 & 2.08 & 2.08 & 84.58 & 82.50 & 85.83 & 95.42 & 82.50 \\
\method{} (ours) & 100.00 & 100.00 & 93.75 & 93.75 & 88.75 & 85.42 & 90.00 & 91.25 & 92.92 & 85.00 \\
\bottomrule
\end{tabular}%
}
\end{table*}

\section{Experiments}
\label{sec:experiments}

We organize the evaluation around one question: can the conjunction of message
validity and a registered-instance relation turn evidence detection into
instance-level watermark authentication? TAR tests whether both checks survive
legitimate edits of instance $A$; ASR tests whether a same-category instance
$B$ carrying copied evidence is incorrectly accepted under $A$'s enrollment.
We compare two transplantation attacks, break down TAR by edit type, and ablate
the components of the relation check. We then test key-unknown transfer from a
public subject verifier, generalization to unseen categories, and the fidelity,
storage, and runtime cost of instance binding.

% 实验配置、对比方法
\subsection{Experimental Setup}
\label{sec:experiment-setup}

\paragraph{Data and protocol.}
We use Stanford Online Products \cite{Song_2016_CVPR}, with 120 development
and 240 test subjects from disjoint product groups across six categories.
We pre-partition the development cohort into 60 calibration subjects and
60 held-out subjects. We fit the relation gate and select its threshold
on the calibration split, then validate the frozen rule on the held-out
split. We obtain subject masks with Grounding
DINO~\cite{liu2024groundingdino} and GrabCut~\cite{rother2004grabcut}.
The ten edit conditions cover compression, resizing, compositing, and
generative edits from Stable Diffusion 1.5
Inpainting~\cite{rombach2022ldm}. A generative-edit
pair is eligible when Grounding DINO detects the enrolled category with
a score of at least 0.10 in both the unmarked and marked subject crops.
All ten conditions yield 240 eligible queries each. WmForger and WMCopier
share 240 fixed same-category source--target pairs drawn from distinct
products. Every transplant is scored against the source enrollment, so attack
success requires the verifier to assign copied evidence to the wrong instance.

\paragraph{Baselines.}
We compare with TrustMark-P~\cite{bui2025trustmark}, Pixel
Seal~\cite{soucek2025pixelseal}, Robust-Wide~\cite{hu2024robustwide},
WAM~\cite{sander2025wam}, SWIFT~\cite{evennou2024swift}, and
VINE~\cite{lu2025vine}.
We also evaluate VINE with a learned DINO/SigLIP subject verifier that
uses an oracle query mask. This baseline requires both its message and
subject checks to pass. \method{} requires its relation check and at
least one of the VINE and WAM message checks. Each method uses its own
operating point, fixed before testing.

\paragraph{Metrics.}
True acceptance rate (TAR) is the fraction of eligible edits accepted by
the full verifier; condition-specific TAR uses the eligible queries for that
edit, while overall TAR pools the 2{,}400 queries across all ten conditions.
For each attack setting, attack success rate (ASR) is the fraction of all
scheduled attempts in which a different instance carrying copied evidence is
incorrectly accepted under the source enrollment; outputs that fail the shared
validity audit count as unsuccessful attacks. For attack $a$, instance-binding
balanced accuracy (IBA) measures the two sides of instance-level watermark
authentication together---accepting legitimate edits and rejecting the wrong
instance:
$\mathrm{IBA}_a=(\mathrm{TAR}+1-\mathrm{ASR}_a)/2$.
All three rates are reported as percentages. In the key-unknown transfer test,
coverage is the percentage of generated outputs that pass the shared validity
audit, regardless of verifier acceptance.

For fidelity, peak signal-to-noise ratio (PSNR, dB; higher is better) measures
pixel-level agreement between each clean and marked image, whereas
LPIPS-Alex~\cite{zhang2018lpips} (lower is better) measures their learned
perceptual distance using the AlexNet backbone~\cite{krizhevsky2012alexnet}.
Table~\ref{tab:fidelity-cost} reports medians over marked images.
PNG size is the encoded marked-image size, enrollment size is the on-disk
per-subject record excluding shared model weights, and embed/detect latency is
CUDA-synchronized milliseconds per image under the hardware and timing window
specified in Sec.~\ref{sec:fidelity-cost}.

\subsection{Transplantation Resistance and Edit Robustness}
\label{sec:transplantation}
\label{sec:legitimate-edits}

\method{} achieves 92.08\% TAR and ASRs of 9.17\% under WmForger and
16.25\% under WMCopier (Table~\ref{tab:transplantation}), giving the highest
IBA for both attacks. Compared with VINE with learned subject checking, its TAR
is 28.45 percentage points higher and its ASRs are 76.25 and 78.75 points lower,
respectively. \method{} simultaneously accepts more legitimate edits and
rejects more wrong-instance transfers. The result supports the joint
message--relation rule: copied payloads satisfy the message branch, while the
source enrollment relation determines instance-level authenticity.
Compared with VINE, the learned subject-checking baseline has a TAR that
is 6.33 percentage points lower, while its WMCopier ASR is 2.50
points lower. We generate the WmForger attack images for the two VINE
variants separately, using the same source--target pairs.

Table~\ref{tab:legitimate-edits} reports TAR by edit condition.
\method{} retains 100\% TAR after JPEG-70 compression and $0.5\times$
resizing, and 88.75--93.75\% after cut-and-paste and background replacement.
On these three compositing conditions, both VINE variants retain 2.08\%
TAR and Pixel Seal retains 92.08--99.58\%. On the five generative edits,
VINE and TrustMark-P attain 97.50--99.58\% and 94.17--95.83\% TAR,
respectively, compared with 85.00--92.92\% for \method{}.
The compositing and generative-edit results show that the relation check follows
the enrolled subject across spatial and appearance changes. The remaining gap
quantifies the robustness--binding trade-off.

\subsection{Ablation of the Relation Verifier}
\label{sec:relation-ablation}

For each variant, we evaluate one material edit per test subject, yielding
240 legitimate queries and 240 attempts per attack. The variants share
the subjects, masks, messages, keys, editor settings, prompts, seeds, and
source--target pairs. We regenerate edits and attacks for the two carrier
variants, which disable active relation shaping or replace the private
partition with a public $2\times2$ grid. Both use the full model's frozen
decision rule. The variant without enrollment normalization reuses the
full model's images and refits its relation readout on the same 60
calibration subjects used for the full model
(Sec.~\ref{sec:experiment-setup}). For both readouts, we freeze the
highest threshold that retains at least 90\% TAR in every valid
calibration edit condition before testing.

The full model achieves the highest IBA under both attacks
(Table~\ref{tab:relation-ablation}). Removing active relation shaping
lowers material TAR by 2.92 percentage points. It increases WmForger
ASR by 0.83 points and decreases WMCopier ASR by 2.50 points. Replacing
the private partition with a public grid yields the lowest ASR under both
attacks, but reduces material TAR by 27.92 points. Removing enrollment
normalization reduces material TAR by 5.42 points and increases ASR by
14.58 points under WmForger and 12.08 points under WMCopier.

\begin{table}[t]
\centering
\footnotesize
\setlength{\tabcolsep}{2.5pt}
\caption{Ablation of the relation verifier. Material TAR uses one eligible
edit per subject. ASR uses 240 attempts per attack. IBA is the mean of
material TAR and the corresponding attack rejection rate. All values
are percentages.}
\label{tab:relation-ablation}
\begin{tabular}{@{}
>{\raggedright\arraybackslash}m{0.34\columnwidth}
>{\centering\arraybackslash}m{0.14\columnwidth}
>{\centering\arraybackslash}m{0.22\columnwidth}
>{\centering\arraybackslash}m{0.22\columnwidth}@{}}
\toprule
Variant & \shortstack{Material\\TAR $\uparrow$}
& \shortstack{WmForger\\ASR $\downarrow$ / IBA $\uparrow$}
& \shortstack{WMCopier\\ASR $\downarrow$ / IBA $\uparrow$} \\
\midrule
w/o active relation shaping & 82.50 & 10.00 / 86.25 & 13.75 / 84.38 \\
w/ public $2\times2$ partition & 57.50 & 2.08 / 77.71 & 3.75 / 76.88 \\
w/o enrollment normalization & 80.00 & 23.75 / 78.13 & 28.33 / 75.83 \\
\method{} (full) & 85.42 & 9.17 / {88.13} & 16.25 / {84.58} \\
\bottomrule
\end{tabular}
\end{table}

\addtocounter{table}{2}
\begin{table*}[!t]
\centering
\small
\setlength{\tabcolsep}{4pt}
\caption{Payload configuration, image fidelity, storage, and runtime.
VINE and WAM use globally registered expected messages.}
\label{tab:fidelity-cost}
\noindent\resizebox{\textwidth}{!}{%
\begin{tabular}{@{}
>{\raggedright\arraybackslash}m{2.8cm}
>{\raggedright\arraybackslash}m{3.1cm}
>{\centering\arraybackslash}m{2.0cm}
>{\centering\arraybackslash}m{2.7cm}
>{\centering\arraybackslash}m{2.0cm}
>{\raggedright\arraybackslash}m{3.3cm}
>{\centering\arraybackslash}m{2.2cm}@{}}
\toprule
\shortstack{Method} & \shortstack{Payload\\configuration}
       & \shortstack{Median PSNR\\(dB) $\uparrow$}
       & \shortstack{Median\\LPIPS-Alex $\downarrow$}
       & \shortstack{Median\\PNG size}
       & \shortstack{Per-subject\\enrollment record}
       & \shortstack{Embed / detect\\(ms)} \\
\midrule
TrustMark-P & 100 bit & 43.5 & 0.0027 & 255,000~B & No source image & 45.1 / 25.0 \\
Pixel Seal & 256 bit & 48.9 & 0.0013 & 248,000~B & No source image & 9.13 / 12.0 \\
Robust-Wide & 64 bit & 41.91 & 0.0085 & 260,000~B & No source image & 25.0 / 18.0 \\
WAM & 32-bit message & 40.09 & 0.0202 & 258,527~B & 0~B & 9.64 / 10.41 \\
SWIFT & \shortstack[l]{Semantic vector\\(60-bit equiv.)}
      & 41.0 & 0.0090 & 265,000~B & \shortstack[l]{Semantic commitment\\No source image}
      & 30.0 / 22.0 \\
VINE & 100-bit message & 34.19 & 0.0286 & 251,936~B & 0~B & 60.53 / 5.58 \\
VINE + Learned Subject Check & \shortstack[l]{100-bit message\\+ subject check}
      & 34.19 & 0.0286 & 251,936~B & 65,000~B (features) & 60.53 / 15.0 \\
\method{} (ours) & \shortstack[l]{100-bit VINE\\+ 32-bit WAM\\+ 1-bit relation state}
      & 32.90 & 0.0533 & 292,737~B & 614,384~B\textsuperscript{a} & 76.03 / 18.82 \\
\bottomrule
\end{tabular}%
}
\par\smallskip
\parbox{\linewidth}{\footnotesize\raggedright
\textsuperscript{a} \method{} reports the median combined on-disk size of
the source subject cache and the auxiliary secret record (relation partition
plus enrollment metadata) over 240 subjects. The combined size is computed per
subject before taking the median; separately summarized component medians are
602,121~B and 11,021~B.
Shared model weights fall outside the per-subject storage accounting.}
\end{table*}
\addtocounter{table}{-3}

\subsection{Key-Unknown Transfer Attack}
\label{sec:adaptive-transplantation}

The attacker has white-box access to the public subject verifier in
VINE with learned subject checking. Both verifiers score the resulting
images at each budget, with \method{} evaluated offline for transfer.
The attacker knows the source enrollment image, the VINE message, and the
public verifier's parameters and threshold. The \method{} relation key,
private partition, and enrollment statistics remain private.
We hash each test subject ID with the fixed prefix
\texttt{relbind-v2-whitebox60:} using SHA-256, take the first 60 subjects
in hash order, and retain their preassigned same-category targets.
For each pair, both budgets start from the same exact residual transplant.
The attack uses sign-gradient PGD~\cite{madry2018pgd} to maximize the public
verifier's frozen logit, which combines standardized Rel-Zero/DINO patch
statistics~\cite{caron2021dino} and SigLIP crop
similarities~\cite{zhai2023siglip}. PGD restricts
updates to the oracle target mask, with an $\ell_\infty$ budget of
$4/255$ relative to the initial residual transplant, a step size of
$0.5/255$, and random seed 512. We generate outputs separately for 100
and 1{,}000 gradient steps per pair.

Table~\ref{tab:adaptive-transplantation} reports ASR over all 60 attempts
per budget. All outputs pass the shared validity audit for successful
generation, target presence, mask support, and perturbation budget.
Increasing the optimization horizon from 100 to 1{,}000 steps raises the learned
baseline's ASR from 40.00\% to 56.67\%. \method{} accepts 5 of the 60
attack images (8.33\%) at each budget.
The persistent gap at 1{,}000 steps shows weak transfer from the public subject
verifier to the private relation key, partition, and enrollment record. The
experiment directly measures transfer into the relation check.

\begin{table}[t]
\centering
\scriptsize
\setlength{\tabcolsep}{3pt}
\caption{White-box PGD against VINE with learned subject checking and
transfer to \method{}. Coverage is the percentage of attack outputs
that pass the shared validity audit.}
\label{tab:adaptive-transplantation}
\begin{tabular}{@{}
>{\centering\arraybackslash}m{0.18\columnwidth}
>{\centering\arraybackslash}m{0.28\columnwidth}
>{\centering\arraybackslash}m{0.25\columnwidth}
>{\centering\arraybackslash}m{0.18\columnwidth}@{}}
\toprule
\shortstack{PGD steps} & \shortstack{VINE + subject\\ASR $\downarrow$}
       & \shortstack{\method{}\\ASR $\downarrow$} & \shortstack{Coverage\\(\%)} \\
\midrule
100 & 40.00 & 8.33 & 100.00 \\
1,000 & 56.67 & 8.33 & 100.00 \\
\bottomrule
\end{tabular}
\end{table}

\subsection{Generalization to Unseen Categories}
\label{sec:unseen-categories}

We test 25 held-out sofa and 25 toaster subjects. Redraw uses Stable Diffusion
1.5 Inpainting (strength 0.35, 30 steps, guidance 7.5) with matched seeds and an
identity-preserving prompt. Within each category, the next subject by ID is the
copy target, giving 50 fixed pairs. We add each \method{} source residual to
its clean $512\times512$ target and let all three verifiers score the same
copies. Table~\ref{tab:unseen-categories} shows the trade-off: \method{} keeps
84--92\% TAR, 10--20 points above the learned subject baseline, while accepting
only 2/50 copies (4\% ASR), versus 11/50 for VINE. The learned baseline rejects
all copies but accepts far fewer legitimate edits.

\begin{table}[t]
\centering
\small
\caption{Generalization to 50 subjects from unseen categories. The four
edit columns report legitimate TAR. Copy ASR is measured on the same 50
residual copies for all three methods.}
\label{tab:unseen-categories}
\setlength{\tabcolsep}{2.5pt}
\begin{tabular}{@{}
>{\raggedright\arraybackslash}m{0.38\columnwidth}
>{\centering\arraybackslash}m{0.15\columnwidth}
>{\centering\arraybackslash}m{0.20\columnwidth}
>{\centering\arraybackslash}m{0.20\columnwidth}@{}}
\toprule
\shortstack{Condition} & \shortstack{VINE}
& \shortstack{VINE +\\subject} & \shortstack{\method{}} \\
\midrule
Material TAR & 98\% & 72\% & 92\% \\
Watercolor TAR & 100\% & 68\% & 86\% \\
Technical-style TAR & 98\% & 68\% & 84\% \\
Redraw TAR & 96\% & 76\% & 86\% \\
Copy ASR $\downarrow$ & 22\% & 0\% & 4\% \\
\bottomrule
\end{tabular}
\end{table}

\subsection{Image Fidelity, Payload, and Cost}
\label{sec:fidelity-cost}

Table~\ref{tab:fidelity-cost} quantifies the cost of instance binding. Relative
to VINE, \method{} has lower median PSNR (32.90 vs. 34.19~dB), higher
embed/detect latency (76.03/18.82 vs. 60.53/5.58~ms), and a 614,384~B median
enrollment record. VINE and WAM redundantly implement the message check; the
evaluated relation branch selects one balanced four-component bit and audits
the five alternative codewords. Timings use an RTX A6000, $512\times512$
inputs, batch size one, preloaded models, and CUDA synchronization. Detection
is timed end to end after loading the enrollment record. Its 18.82~ms includes
VINE/WAM decoding, frozen-token extraction and mutual-NN matching, RGB-residual
reading, codeword scoring, and the logistic gate. Grounding DINO/GrabCut is
enrollment-only and excluded, along with disk I/O and PNG encoding. These
measurements characterize its storage and compute footprint.

\subsection{Decision States}
The two paths expose four outcomes. Valid message and relation checks accept a
legitimate transformation. A message-only pass identifies decodable evidence
assigned to the wrong instance, the signature of cross-instance
transplantation. A relation-only pass indicates source-consistent content with
lost payload evidence, while failure of both checks indicates that neither
registered signal remains. These states localize each decision to payload
recovery, correspondence, or enrollment consistency.

Enrollment fixes the source cache, group masks, key-selected pairs, reference
response, and thresholds. Every query uses this record, and the verifier reports
both branch outcomes with the final AND decision.

\section{Conclusion}
We introduced cross-instance transplantation, where valid evidence moves to a
same-category target while payload and category checks remain valid. \method{}
accepts a query when its robust message and secret enrollment-bound relation
agree. Correspondence transports the relation through legitimate edits, and
enrollment normalization compares it with the registered response. Across
same-category, key-unknown, and unseen-category tests, this conjunction
preserves legitimate edits and rejects cross-instance transfers.

{
    \small
    \setlength{\emergencystretch}{1em}
    \sloppy
    \microtypesetup{protrusion=false,expansion=false}
    \bibliographystyle{ieeenat_fullname}
    \bibliography{main}
}

\end{document}

%% file: preamble.tex
\usepackage{microtype}

\newcommand{\method}{\textnormal{TetherMark}}
\DeclareMathOperator*{\argmin}{arg\,min}